\documentclass[preprint2]{aastex}   % for twocolumn
\usepackage{epsfig,graphicx}

\def\chandra{{\it Chandra\/}\/}

\def\eohone{E0102.2$-$7219\,}
\def\dem{DEM~L71\,}
\def\einstein{{\it Einstein\,}}

\def\asca{{\it ASCA\,}}
\def\kms{km s$^{-1}$\,}

\def\myarcsec{$\mskip1mu^{\prime\prime}\mskip-7mu\mskip2mu$}

\def\memp{$m_{\rm e}/m_{\rm p}$\,}
\def\gmemp{$g_{0} = m_{\rm e}/m_{\rm p}$\,}
\def\te{$T_{\rm e}$\,}
\def\tp{$T_{\rm p}$\,}
\def\etal{et~al.\,}
\def\net{$n_{\rm e}t$\,}

\received{2002 November 22}
\begin{document}

\title{The Physics of Supernova Remnant Blast Waves. II. Electron-Ion Equilibration in \dem in the Large Magellanic Cloud}
\author{
Cara E. Rakowski\altaffilmark{1},
Parviz Ghavamian\altaffilmark{1},\and
John P. Hughes\altaffilmark{1}
}
\affil{Department of Physics and Astronomy, Rutgers The State
University of New Jersey, 136 Frelinghuysen Road, Piscataway NJ 08854-8019}
\altaffiltext{1}{E-mail: rakowski@physics.rutgers.edu, parviz@physics.rutgers.edu, 
jph@physics.rutgers.edu
}

\begin{abstract}

We present analysis and modeling of X-ray spectra from the blast wave
shock of \dem in the Large Magellanic Cloud.  This remnant exhibits
widespread Balmer-dominated emission characteristic of nonradiative
shocks in partially neutral gas.  We have used both {\it Chandra}
ACIS-S data and optical Fabry-Perot spectra of the blast wave to
measure the electron and proton temperatures, respectively.  In
principle, when combined, these measurements can determine the degree
of electron-ion temperature equilibration ($g_{0} \equiv T_{\rm
e}/T_{\rm p}$) immediately behind the shock front.  In our X-ray
analysis we fit {\it Chandra} spectra of three nested regions behind
the blast wave under three different scenarios: (1) a planar,
initially unequilibrated shock (\gmemp), where the downstream electron
and proton temperatures equilibrate through Coulomb collisions, (2) a
planar, immediately equilibrated shock ($g_{0} = 1$) and (3) a spherical,
equilibrated shock under Sedov evolution.  Using independent
measurements of \te and \tp we find that the X-ray spectra from the
fastest blast wave locations ($V_{s} \sim 700-1000$ \kms) are
consistent with little or no equilibration at the shock front and are
inconsistent with full equilibration.  In contrast, spectra from
regions showing slower blast wave speeds ($V_{s} \sim 400 -600$ \kms)
allow full equilibration but exclude zero equilibration.  In order to
directly constrain the initial equilibration, we incorporated our
knowledge of the proton temperatures into our X-ray models to build
planar shock models that allowed for a variable $g_{0}$.  This model
confirmed and strengthened the above results.  Specifically, we found
that X-ray spectra from an intermediate velocity shock ($V_{s} \sim
800$ \kms) were consistent with intermediate equilibration, excluding
both \gmemp and $g_{0} = 1$ at greater than 1$\sigma$.  Overall, our
results support the picture of decreasing electron-ion equilibration
with increasing shock speed found from previous studies of optical
spectra in other Balmer-dominated supernova remnants.

\end{abstract}


\section{Introduction}

The outer blast wave of \dem\ provides a valuable laboratory to
address the question of electron-ion equilibration at collisionless
shocks.  These shocks have been traditionally explained by their
Rankine-Hugoniot jump conditions (e.g., Spitzer 1978), which quantify
the partition of energy into thermal and bulk kinetic forms.  However,
for collisionless shocks the actual heating is not straightforward
because at extremely low densities, the shock front is thinner than
the mean free path for collisions. Therefore, the heating of particles
at the shock-front must be produced by something else, generally
assumed to be plasma waves (for a review see for instance, Draine \&
McKee 1993; Laming 2000).  Since the heating is not collisional
there is no guarantee that it will produce a thermal distribution, and
in fact both simulations and in-situ observations of solar-wind shocks
suggest that it does not. The plasma heating mechanisms, to both
thermal and cosmic-ray energies, are as yet poorly understood. These
collisionless shocks are ubiquitous in astrophysical situations, but
are only spatially resolved in interplanetary shocks (at lower speeds)
and SNRs. The partition of energy into thermal and non-thermal
populations at collisionless shocks has broad reaching implications
for the dynamics of the ISM, shocks from merging galaxies, and hot gas
in clusters of galaxies. Our previous work on \eohone\ studied the
partition of energy into thermal electrons versus cosmic rays (Hughes,
Rakowski, \& Decourchelle 2000). Here we examine the partition of
energy between the two main thermal populations, electrons and
protons.

In a small sub-sample of supernova remnants (SNRs), the X-ray and
radio emission from the blast wave is accompanied by optical
(H$\alpha$, Balmer) line emission.  The line profiles of these
Balmer-dominated spectra provide a direct measurement of the proton
temperature at the shock front, thus also placing limits on the shock
velocity.  The optical spectra of Balmer-dominated SNRs are produced
by collisional excitation when partially neutral interstellar hydrogen
is overrun by the blast wave (Chevalier \& Raymond 1978). Like other
non-radiative shocks, post-shock cooling losses are negligible for
Balmer-dominated remnants.  Each Balmer emission line consists of a
narrow and a broad velocity component (Chevalier, Kirshner \& Raymond
1980).  The narrow component is produced when cold H~I, overrun by the
shock, is collisionally excited by electrons and protons before being
ionized. The cold neutral hydrogen can also charge exchange with fast
post-shock protons, producing fast neutrals. These can then be
collisionally excited, producing the broad velocity component
(Chevalier et al.~1980).  The Balmer emission arises in a thin
($\lesssim 10^{15}$ cm) ionization zone; therefore, the width of a
broad Balmer line yields the proton temperature immediately behind the
shock and hence also limits the range of possible shock velocities
(Chevalier et al.~1980, Smith et al.~1991).  Combining this
information with the broad-to-narrow flux ratio allows us to estimate
the degree of electron-ion temperature equilibration, $g_{0} \equiv
(T_{\rm e}/T_{\rm p})_{0}$, due to plasma processes at the shock front
(Tycho, SN 1006 \& RCW 86; Ghavamian \etal\ ~2001).  

A more straightforward method of estimating the initial temperature
equilibration in a collisionless shock is to combine the proton
temperature estimated from the FWHM of the broad Balmer line with the
electron temperature measured from X-ray spectra.  For this purpose,
the Large Magellanic Cloud remnant \dem\, is particularly appropriate
because it is completely encircled by Balmer-dominated filaments
(Tuohy \etal\ ~1982).  We have utilized Fabry-Perot imaging
spectroscopy to measure the width of the H$\alpha$ broad component
along most of the rim of \dem (Ghavamian \etal\ ~2003; hereafter
GRHW03).  Here we report results comparing \te from our {\it Chandra}
analysis with \tp from our Fabry-Perot analysis to infer the value of
$g_{0}$ for multiple apertures around the blast wave of \dem.

Davies, Elliot, \& Meaburn (1976) first identified DEM L71 as a SNR
candidate in an optical survey of the Large Magellanic Cloud (LMC)
because of its shell-like morphology that extended 83\myarcsec\ by
60\myarcsec\ in diameter.  The detection of X-ray emission from DEM
L71 in the \einstein\ survey of the LMC by Long, Helfand and Grabelsky
(1981), confirmed its nature as a SNR. DEM L71 is extremely faint in
the radio band, where it has only been detected at 843 GHz with a flux
of 10 mJy by Mills et~al.~(1984). In optical follow-up, Tuohy
et~al.~(1982) detected filamentary H$\alpha$ emission but only a few
faint knots of [O~III] and [S II], and thus categorized it as a
Balmer-dominated SNR.  Smith et~al.~(1991) included \dem\ in their
study of 6 Balmer-dominated remnants. Using longslit spectroscopy of
the brightest portions of the shock in \dem\, they estimated shock
velocities of 300$-$800~km~s$^{-1}$ from the H$\alpha$ broad component
widths. In a recent \asca study of LMC SNRs, Hughes, Hayashi \& Koyama
(1998) fit the X-ray spectrum of \dem and showed that the data were
well described by a non-equilibrium ionization (NEI) Sedov model,
particularly for the case of only minimal initial equilibration
between electrons and ions.

Further investigations into the nature of \dem\ utilizing the
\chandra\ observation are given in two companion papers. GRHW03
discusses the details of the Fabry-Perot analysis of the blast wave,
and implications for the evolutionary state of the remnant. In Hughes
et~al.~(2003) we investigated general properties of the remnant and
the composition of the ejecta. \dem\ exhibits a prominent double shock
morphology with a clear spatial separation between the reverse shock
(ejecta) and blast wave (ISM). The mass ($\sim 1.5 M_\odot$) and
composition of the ejecta (Fe and Si-rich), strengthen the case for a
Type Ia SN origin for \dem .

\section{Observations and Data Reduction}
\subsection{RFP optical observations}

Optical spectra of \dem\ were obtained using the Rutgers Fabry-Perot
(RFP) imaging spectrometer as described in GRHW03. For completeness we
summarize the relevant analysis here.  The 14 RFP scans, centered on
the wavelength of $H\alpha$ at the systemic velocity of the LMC,
provide essentially monochromatic images of the SNR. H$\alpha$ line
emission spectra can then be extracted from any position on the
remnant. We defined 16 apertures around the blast wave, which
contained approximately equal numbers of counts and avoided any stars
or radiative filaments. Fits of the H$\alpha$ line profile determined
the broad component widths and broad-to-narrow flux ratios for each
aperture. For the purpose of comparison to the X-ray results, larger
regions were required to obtain reasonable statistics in the \chandra\
X-ray spectra. Adjacent regions with consistent broad component widths
and broad-to-narrow ratios were combined to provide five final optical
apertures (Figure~1, Table~1). The H$\alpha$ line profiles from these
apertures were then refitted.  The proton temperature, \tp , and shock
velocity, $V_{s}$, were calculated from the measured broad component
width for the full range of $g_{0}$ (GRHW03).  The values computed for
the extrema of \gmemp and $g_{0}$=1 are reprinted here in Table~2.

\subsection{\chandra\ X-ray observations}

We observed \dem\ using the back-side-illuminated chip (S3) of the
Advanced CCD Imaging Spectrometer (ACIS-S) in full-frame timed
exposure mode starting on 2000 January 04 for 45.4 ks (OBSID 775).  At
this time, the ACIS focal plane temperature was $-$110$^\circ$ C.  The
starting point of the reduction was the events lists from the revision
2 level of the standard processing.  The CCD events were corrected for
charge transfer inefficiency (CTI) using software developed at Penn
State University (Townsley et~al.~2000).  Once these corrections are
applied the entire chip can be modeled with a single response
function.  Subsequently, the event file was screened for grade
(retaining only values 0, 2-4, or 6), bad pixels, high background
times and times with incorrect aspect.  The final filtered events file
corresponds to a livetime-corrected exposure of 34.5 ks.

Using the optical image of \dem\ as a template, we verified that the
\chandra\ X-ray positions are accurate to
$\sim$0.5$^{\prime\prime}$. This method was used because none of the
16 X-ray point sources detected on the S3 chip were coincident with an
optical counterpart from the USNO-A2.0 star catalog.

The spectrum of the blast wave is dominated by emission below 0.8
keV. Therefore, this soft band was used to define the blast wave
regions matched to each optical aperture.  We adaptively smoothed the
soft band image using IMSMOOTH from the {\it Chandra Interactive
Analysis of Observations} (CIAO) package (with a minimum
signal-to-noise ratio of 4), then chose a contour level of 0.8 counts
per pixel to define the edge of the blast wave. Three polygonal
regions, each four pixels wide ($\sim 2^{\prime\prime}$), were defined
inward of the outer contour towards the local center of curvature
(Figure 2). Blast wave spectra were extracted from the filtered events
file using these nested regions. The three succeeding regions were
constructed in order to determine not only the blast wave electron
temperature but also its downstream evolution.  The largest
temperature gradient is expected close to the current shock
front. This and the fact that shortly behind our defined regions other
shock fronts and the interior (reverse shock) emission become evident,
limited the distance over which we could extract blast wave spectra.
In turn, this limited the number of nested regions to three because we
required the extraction regions to be wider than the \chandra\
point-spread-function as well as wide enough to obtain sufficient
statistics.  Note that the length of each X-ray extraction region was
defined to match the full extent of the corresponding optical
apertures.

Separate ancillary response files (ARFs) for each spectral extraction
region around the blast wave were generated using standard CIAO
procedures. Although these were retained in the final fits, we note
here that no significant differences between the ARFs were found.  For
all spectra, background was taken from a large annulus outside the
remnant covering radii 1$^{\prime}$--2$^{\prime}$.  We did not attempt
to account for the recently discovered degradation in the low energy
quantum efficiency of the CCD.\footnote{see the calibration area of
the \chandra\ website for the announcement and details:
\texttt{http://cxc.harvard.edu/cal/Acis/Cal\_prods/qeDeg/index.html} }
Instead we compensate for this effect by allowing the column density
in our fits to be a free parameter.

\section{X-ray modeling}

\subsection{Theory}

To measure the electron temperature at the blast wave, one has to make
some assumptions about the evolution of both the temperature and
ionization state downstream. These assumptions may bias the measured
electron temperature and hence the final derived value $g_{0}$. To
address this concern we chose to investigate three very different
nonequilibrium ionization (NEI) evolutionary models for the blast
wave: (1) a fully equilibrated ($g_{0} = 1$), constant velocity,
planar shock model, (2) an initially unequilibrated (\gmemp\ ) planar
shock model with post shock equilibration on Coulomb collisional
timescales (Spitzer 1978), and (3) a fully equilibrated ($g_{0} = 1$)
Sedov model.  In all three models there are only two important
dynamical parameters, the current average temperature at the shock
front, $T_{ave,s}$, and the final ionization timescale, \net\, of the
innermost shock region.  All other temperatures and timescales are
constrained to follow the appropriate evolution. The temperatures
within a parcel of gas that may evolve as a function of time are the
electron temperature \te, the proton temperature \tp , and their
number-density-weighted average temperature, $T_{ave}$. Note that
throughout this paper, "average temperature" is used to denote the
number-density-weighted average over the particles, and not a spatial
or temporal average. All ions were considered to have the same
temperature as the protons for the sake of the average temperature, a
small effect for the metal-poor LMC interstellar medium.

Planar shocks are a reasonable approximation for (1) straight or small
shock segments where the curvature is unimportant for the density
profile, and (2) time periods that are short enough that the
deceleration of the blast wave due to the accumulation of swept up
material is negligible. The planar model has a constant velocity and
hence a constant average temperature behind the shock.  Initially, we
chose to model the two extreme cases for the initial degree of
electron-ion equilibration, $g_{0} = 1$ and \gmemp. For $g_{0} = 1$,
all temperatures (\te, $T_{ave,s}$, and \tp) are equal and constant
throughout the shock and only the ionization timescale varies behind
the shock. For \gmemp, \te and \tp slowly equilibrate to $T_{ave,s}$
via Coulomb collisions (Spitzer 1978).  We use the expressions for the
time variation of temperature from Itoh (1978) and Cox \& Anderson
(1982) to describe the variation of \te as a function of $n_{\rm e}t$
behind the shock.  Note that while this formulation was derived for
Sedov evolution, it is actually equally appropriate for planar shocks.
For examples of the evolution of planar shocks see section~5.

Considering the advanced age of DEM L71 ($\sim$4400 years, GRHW03) it
was important to include a Sedov solution model.  Even under full
temperature equilibration, the Sedov solution implies a spatially
varying electron temperature behind the shock, which could mimic the
temperature variation from Coulomb equilibration in a $g_{0} = m_{\rm
e}/m_{\rm p}$ planar model.  The Sedov solution used here assumes a
spherical geometry and uniform preshock medium.  For simplicity, we
only modeled the case of full electron-ion equilibration, i.e., \te =
$T_{ave}$ = \tp. There are two mechanisms that determine the evolution
of $T_{ave}$ and $n_{\rm e}t$. First, the shock decelerates as it
sweeps up interstellar gas, causing $T_{ave,s}$ to decrease with time.
Second, the volume adiabatically expands as the shocked material moves
out at $(3/4) V_{s}$, i.e., the interior density and temperature
decrease.  Parcels of gas that are shocked at different radii,
$r_{i}$, and times $t_{i}$ started off with different $T_{ave,s}$ and
$V_{s}$. Hence each parcel will follow a unique temperature and
density evolution.

The beauty of the Sedov solution lies in its self-similarity.  All
shock quantities can be related by dimensional analysis, and the
interior profiles can be expressed in terms of dimensionless reduced
variables.  Hence it is possible to construct a generic set of reduced
temperature and density histories as a function of current fractional
radius (see Figures 3 and 4). We start with the Sedov solutions for
the shock quantities and the interior profiles, herein denoted as the
shock and interior solutions respectively (see Sedov 1959).  The shock
solutions can be written to relate the current radius ($R_{s}$),
temperature ($T_{ave, s}$), velocity ($V_{s}$), and the swept up mass
($M_{SU}$) to the explosion energy, the hydrogen number density of the
ambient medium ($n_{0}$), and the time since the explosion ($t$).
Following Cox \& Franco~(1981), we express the interior solutions in
terms of the dimensionless quantity $\beta$ which ranges from 0.8 to 1
from the center to the edge of the blast wave and finely samples the
outermost radii.\footnote{There is a minor typographical error in the
Cox \& Franco~(1981) definition of $\beta$: the coefficient three
tenths should be inverted to ten thirds, in order to match the
solutions given by Sedov (1959).} This variable is also convenient
because it only depends on nondimensional combinations of the
parameters in the problem.  These interior solutions describe the
profiles of reduced variables, the ratios of physical quantities
relative to the immediate post-shock values: radius ($r$), density
($x$), pressure ($y$), initial radius ($r_{i}$), and the mass enclosed
($\mu$) within that $\beta$. The reduced temperature is just the ratio
of the reduced pressure to the reduced density. For clarity, we note
that for a given parcel of gas there are three times of importance in
the formulation of the temperature and density histories: the time at
which the parcel was shocked, $t_{i}$; the time at which the evolution
is being evaluated, $t$; and the current epoch, $t_{c}$.

To derive the density and temperature histories of a parcel of gas as
a function of $t$, we need to connect the shock and interior
solutions.  We make use of the fact that the mass enclosed behind the
radius of a given parcel of gas is constant, i.e., that portions of
the post-shock material never overtake the gas ahead of them. This
allows us to match the equations for the swept up mass at the time the
parcel was shocked, $M_{SU}(t_{i})/M_{SU}(t)$ with those for the
fraction of the total swept-up mass that is enclosed within a given
interior radius, $\mu (\beta) = M_{enclosed}/M_{SU}$ to get
$\beta(t_{i}/t)$.  This $\beta$ history defines where any given parcel
was within the remnant as a function of time, as well as its density
and temperature relative to the shock values.  We can then combine
this with the shock solutions which relate the shock quantities at $t$
to the current shock quantities. Thus we obtain final solutions for
the ratio of the evolving $T_{\rm ave}$ and $n_{\rm e}t$ of each
parcel to the current shock values in terms of the ratio of $t$ to
$t_{c}$.  These generic temperature and ionization timescale histories
can be multiplied by any chosen time, density and temperature to give
specific solutions.

There are a number of important features of the Sedov solution that
can be seen in the generic temperature and ionization timescale
histories shown in figures 3 and 4.  The three curves shown correspond
to the current fractional radii of the three X-ray spectral extraction
regions behind the shock in \dem.  They represent the evolution of
parcels of gas whose final radii are at the innermost edges of these
three regions.  Note the steep decrease in temperature both in time
and current radius.  Material that now lies at $0.85 r_{s}$ was
shocked at about one-fifth the lifetime of the remnant when the shock
was almost six times as hot. Even the current temperatures across the
three spatial zones vary by more than a factor of two.  This leads to
a spatial gradient in the electron temperature for the Sedov model
that is far larger than the variation in temperature across regions
for the minimally equilibrated (\gmemp ) planar shock model, which is
at most 40\% (see Table 5).

The ionization timescale of the Sedov model also has a very different
evolution than in the planar case.  Adiabatic expansion of the post
shock gas causes both a decrease in temperature as well as a decrease
in the density, reducing the rate of increase of $n_{\rm e}t$. The
global maximum value of $n_{\rm e}t$ is reached for parcels that are
now at $\sim 0.9 R_{s}$.  However, the temperature of the parcel which
is currently at 0.85$r_{s}$ was more than three times hotter than the
current $T_{ave,s}$ for almost half of its evolution.  Since the
ionization rates depend on the temperature, the final $n_{\rm e}t$ by
itself is not a very accurate measurement of the ionization state
profile behind the shock. This complicates a direct comparison of the
$n_{\rm e}t$ values between the planar and Sedov models.

\subsection{NEI implementation}

Other authors have implemented non-equilibrium ionization in Sedov and
planar geometries (Shull 1982; Hamilton et~al.~1983; Kaastra \& Jansen
1993; Hughes et~al.~1998; Borkowski, Lyerly \& Reynolds 2001) but for
the sake of flexibility and the specific needs of the selected regions
in \dem , we chose to modify our own NEI code (Hughes \& Singh 1994).
In all cases we construct one evolving shock model that is then used
to fit the three nested regions simultaneously.

For both the \gmemp and $g_{0} = 1$ planar models, we divided the
post-shock region into 200 logarithmically spaced regions in \net .
Since the average shock temperature is constant in the planar model,
all parcels of gas will undergo the same electron temperature and
ionization state history. At a constant shock velocity, the \net\
steps represent both the time since the material was shocked and the
distance behind the shock.  This allows us to calculate the emission
at each timestep and sum them to find the total emission from the
shock.  For each step we calculated the ionization state and emission
using the electron temperature at that time determined by the solution
of Itoh (1978) and advanced the ionization state of the material
during that \net\ interval as in Hughes \& Singh (1994). The final
ionization state from each step was then used as the initial
ionization state for the following interval.  The emissions were
summed, weighting by $\Delta n_{\rm e}t$, which for a planar shock is
also the width perpendicular to the shock. For the purpose of fitting
the nested \dem\ blast wave spectra, each model was cut into 3
sections linearly spaced in \net.  We assumed that the current blast
wave lies at the outer edge of the outermost aperture.  However, the
fitted temperatures were not particularly sensitive to the exact
choice of blast wave location. Finally, we included a simple spherical
projection of the three modeled spatial regions to account for the
contribution of the outer emission zones to the inner spectral
regions. The projection in this case was done only for the three large
data extraction regions and not the individual \net\ intervals, so
that the different values of $R_{s}$ for each aperture could be
accounted for during the fits.

For the Sedov NEI model, first the generic histories were scaled to a
given current shock temperature and product of the ambient density
with the remnant age.  The ionization state of the material at every
radius (parcel) was calculated separately using its $T_{\rm e}$ and
$n_{\rm e}t$ history, then the emission was computed for the current
temperature of that parcel. A grid of 100 points in $\beta$ (over its
full allowed range of 0.8 to 1.0) was used for both the time evolution
histories and the emission as a function of radius.  A single set of
nominal fractional radii (0.85, 0.89, 0.95, calculated for an 8.8~pc
current radius) was used for all regions in DEM L71, despite their
slightly different radii (see GRHW03 figures 7 and 8), so that a
common grid of models could be applied to all spectra.  At these
nominal fractional radii spherical projection of every individual
segment in $\beta$ was used to weigh the contribution of any given
parcel to each of the nested regions behind the shock.  Note that the
Sedov model is inherently spherical, in contrast to the case described
in the preceding paragraph where we were simply adding a spherical
projection to an otherwise planar model.

\section{X-ray spectral fits}

In the preceding section, we introduced the dynamical variables,
$T_{ave,s}$ and \net\ that we wish to measure with our X-ray
fits. However, the spectra also depend on the composition of the gas
and the absorbing column density to the SNR. Ideally, one would like
to allow the abundance and column density parameters to be free for
each region to allow for variations in the composition of the
interstellar medium around the rim of the remnant, and any local
absorption. Furthermore, assuming a particular abundance set may bias
the temperature measurements, because they are not independent
variables. For example, choosing lower abundances will often bias the
fits towards higher temperatures. Unfortunately our preliminary fits
revealed that the current statistics are insufficient to constrain
these additional parameters.

For both planar and Sedov fits the abundances of O, Ne and Fe were at
first allowed to be free, with all other species fixed at the LMC
abundances of Hughes et~al.~(1998). The absorbing column density,
$N_{\mathrm H}$, was also free in these preliminary fits. The
abundances included in our column density parameter were kept at solar
values. Since the total column densities we found were similar to the
minimum absorbing gas column in the Galaxy along the line of sight to
the LMC ($5-6 \times 10^{20}$ atoms~cm$^{-2}$, Heiles \& Cleary 1979),
no additional LMC component was deemed necessary.  From these
preliminary fits we determined the average abundances and column
density across all regions and all models.  Most regions did not have
high enough statistics to constrain the individual
abundances. Furthermore, the F-test showed that freeing the abundance
and $N_{\mathrm H}$ parameters did not significantly improve the fit
over using the average values.  Hence a common set of abundances for
all models and regions was used in order to allow a more
straightforward comparison both between different regions and for a
single region under different models.  In the final fits we fixed the
abundances and column density to three sets: the average values and
the average values plus or minus their root mean square (RMS)
deviations.  The final values are listed in Table 3.  They are all
reasonable for the interstellar medium in the LMC compared to the
abundances of Hughes et~al.~(1998).  We fitted the spectra using the
RMS deviations in the column density and abundances in order to
incorporate the uncertainty in the temperature and timescale due to
(1) the uncertainty in our estimate of the average abundances and
column and (2) possible variations in abundances and column density
around the outer rim of the remnant. This estimate of the errors is
quite conservative since the RMS errors are much larger than the
errors in any given preliminary fit and include any variations due to
the choice of model.

The uncertainty in the abundance and column density parameters
prevented us from distinguishing the three shock models on the basis
of their $\chi ^2$ values. Not only was every model the ``best''
choice for some region based on $\chi ^2$, but the variation in $\chi
^2$ across models was always smaller than across the RMS deviations
in abundance and column density (not shown).

The nested spectra for each of the five apertures are shown in Figure
5, plotted against the \gmemp , planar model for comparison. The
best-fit parameters of the final fits for all regions and all shock
models are given in Table 4. The average shock temperatures predicted
by the three shock models are very different because the X-ray
emission depends primarily on the current electron
temperature. Roughly speaking then, the values for $T_{ave,s}$ are
just those that give the correct $T_{\rm e}$ for the ``dominant''
region in any given aperture under the assumptions of that shock
model. The implied average shock temperatures are useful, however, for
comparison with the optically determined proton temperatures.

Under the assumption of full equilibration, $g_{0}=1$, the electron,
proton and average temperatures are all equal at the shock. Hence if
either the planar or the Sedov $g_{0}=1$ models are correct the
optically measured $T_{\rm p}$ should match the X-ray fitted
$T_{ave,s}$. This is the case only for region X5, the
blast wave region with the slowest shock speed. For all other regions
$T_{\rm p}$ is much greater than the $T_{ave,s}$ values for full
equilibration, implying that the electron temperature at the shock is lower
than the proton temperature. The \gmemp\ planar fits confirm this
result. Under the \gmemp assumption, $T_{\rm p} \simeq (n_{\rm e} +
n_{\rm p})T_{ave}/n_{\rm p} $ or $T_{\rm p} \simeq (2.3/1.1) T_{ave}$. 
For all regions except X5, the X-ray implied $T_{\rm p}$ is
consistent with the H$\alpha$ proton temperature within the 1$\sigma$
errors.
This agreement occurs on the low side of our allowed $T_{ave,s}$ range
hinting that partial equilibration may be preferred.

All three NEI models indicate the same general result: by comparing
the average shock temperatures to the H$\alpha$ proton temperatures,
we can already exclude the case of full electron-ion equilibration in
four out of five apertures, independent of the global model. In
contrast, region X5, which has the slowest shock speed, is consistent
with $g_{0} = 1$ but not with $g_{0} = m_{\rm e}/m_{\rm p}$ under the
assumption of a constant velocity shock, and is marginally consistent
with $g_{0} = 1$ under the Sedov model.

In addition to a comparison with the optically measured proton
temperature, the parameters of the X-ray fit also imply a certain
evolution of the electron temperature and ionization state.  Table 5
lists the implied values of $T_{\rm e}$ and \net\ for each of the
nested regions behind the shock in the three NEI models. The different
model assumptions lead to contrasting spatial gradients in the
electron temperatures. The ionization timescale values also differ
strongly from model to model, but this can be understood in terms of
their underlying assumptions.  For the \gmemp constant velocity model
the fitted \net\ values are larger by factors of 10\% to 100\% than
those for the $g_{0}=1$ constant velocity model.  Early on the
electron temperatures were lower and hence ionization progressed less
rapidly. The fits therefore require a larger \net\ value to reach the
same ionization state. For the fully equilibrated Sedov model, the
final ionization timescales are much lower than the $g_{0}=1$ planar
case, because earlier the material was much hotter than now, so the
ionization occurred more rapidly.

Deeper observations, with sufficient statistics to
constrain the electron temperature and ionization state in each of the
nested regions separately, would clearly be able to discriminate
between the three models given here, based on their strong differences
in the implied evolutions.

\section{Estimates of the Initial Electron-Ion Temperature Equilibration}

In the previous section we showed that our independent measurements of
the electron and proton temperatures led to a consistent picture of
the initial temperature equilibration at each blast wave region
regardless of which NEI model was used.  Here we will assume a
constant velocity shock and present an estimate of $g_{0}$ from the
independent $T_{\rm e}$ and $T_{\rm p}$ measurements, and a direct
determination of $g_{0}$ using the proton temperature to constrain our
X-ray models.

In figure 6 we plot the electron temperature as a function of
ionization timescale for three types of constant velocity shock
models, using region X1 for illustration.  The diamonds denote the
electron temperatures and ionization timescales implied by the fitted
values of $T_{ave,s}$ and \net\ for the fully equilibrated X-ray
model, while the crosses are those for the \gmemp model. The curves
plot the Coulomb evolution of the electron temperature given the
best-fit value of the proton temperature, and the full range of
possible $g_{0}$ values.  All curves eventually equilibrate (flatten
out) to the average temperature.  The average temperature is not the
same for all $g_{0}$ values, because the initial proton temperature is
kept constant: $T_{ave,s}$ will be lower if the electrons are not
fully heated initially.  Hence, the top curve represents full initial
temperature equilibration, where $T_{\rm e}$, $T_{\rm p}$, and
$T_{ave}$ are equal and remain constant, while the bottom curve
represents the minimal initial heating case, \gmemp .

From Figure 6 it is evident that the electron and proton temperatures
are indeed consistent under a planar model with some degree of initial
electron heating followed by Coulomb equilibration.  The actual
estimate of $g_{0}$ will depend on which X-ray model is used to derive
the electron temperature to compare to the proton temperature, as well
as which region behind the shock is chosen.  However, the relative
electron temperatures and the ionization timescale differences across
the three regions are simply fixed by the model assumptions. 
For a first estimate of $g_{0}$, we chose to use the \gmemp planar
model to constrain $T_{\rm e}$ at the outermost region (i.e., shortest
ionization timescale, $n_{\rm e}t$).  We compared this to the electron
temperature at that timescale implied by the proton temperature as a
function of $g_{0}$.  These estimates are listed in Table 6. The
errors include those due to $T_{\rm p}$, $T_{\rm e}$, and abundances
summed in quadrature, which were all of approximately equal
importance.  As seen before from the comparison between $T_{ave,s}$
and $T_{\rm p}$, only X5, the aperture with the slowest shock speed,
is consistent with $g_{0}=1$. All others are consistent with \gmemp
. The large errors in both $T_{\rm p}$ and $T_{\rm e}$ when added in
quadrature lead to large errors in $g_{0}$.

Given that our independent measurements have confirmed that the blast
wave temperatures are consistent with a planar shock with some degree
of initial electron-ion equilibration, it is now possible to constrain
$g_{0}$ by incorporating our knowledge of the proton temperature into
the X-ray model fits.  We expect this to greatly reduce the errors in
our determination of $g_{0}$ for two reasons. Firstly, $T_{\rm e}$,
$T_{\rm p}$, \net\, and the abundances are all correlated variables
and hence summing their errors in quadrature over-predicts
the actual error. Secondly, our three nested regions give us some
constraint not only on the average electron temperature but also its
spatial gradient. Allowing $g_{0}$ to vary gives us access to
different temperature evolutions, beyond the three extreme cases
tested before.

We adapted the planar model to consider a single $T_{\rm p}$ and allow
for different values of $g_{0}$ from \memp to 1.  For each aperture we
built a grid of models for the best-fit $T_{\rm p}$ as well as its
$1\sigma$ error limits. The only free parameters are then $g_{0}$ and
\net . The average shock temperature and velocity are dependent on
$T_{\rm p}$ and $g_{0}$ (see the above discussion of Figure 6). For
any given $g_{0}$, the electron temperature varies with distance
behind the shock as the downstream gas undergoes Coulomb collisions.
For consistency, the abundances and column density values were fixed
to the same values used in the previous fits.

The fits to the variable $g_{0}$ model are as good as the previous
three. The resultant $g_{0}$ values (listed in Table 7) are consistent
with the previous cruder estimates but with tighter constraints.  In
many regions \net\ and $T_{\rm p}$ work together to give very similar
$g_{0}$ values within the $1\sigma$ range on $T_{\rm p}$. One
aperture, X4, is found to be of intermediate equilibration, excluding
both \gmemp and $g_{0}=1$ (at the $1\sigma$ level).

What do these $g_{0}$ values imply for the relationship between
velocity and initial equilibration in collisionless shocks?  In Figure
7 we compare our DEM~L71 results on $g_{0}$ as a function of $V_{s}$
with previous work on other remnants using optical observations of the
H$\alpha$ line alone (Ghavamian et~al.~2001). {\it The results are
consistent and indicate a decreasing level of equilibration with
increasing shock speed.} The method presented in the current paper is
considerably different than the previous work modeling the H$\alpha$
broad-to-narrow ratios. H$\alpha$ emission is only produced extremely
close to the shock front, so the $g_{0}$ values modeled by this method
truly reflect the initial equilibration, modulo assumptions about the
preshock ionization fraction.  In contrast the X-ray emitting region
is extended. We measure the electron temperature some distance
downstream of the shock and then infer the initial equilibration,
assuming that only Coulomb collisions have equilibrated the
temperatures thereafter.  That these two different methods lead to
consistent $g_{0}$ values as a function of shock velocity validates
both methods. Furthermore it suggests that there are no significant
additional heating mechanisms beyond Coulomb collisions that operate
downstream of the immediate shock zone. The real test, of course, will
come when both methods can be applied to the same shock
front. Unfortunately this was not possible here, because the
broad-to-narrow ratios were anomalously low, most likely due to
precursor contamination of the narrow component of the H$\alpha$
emission (see GRHW03 for more details).

\section{Summary \& Concluding Remarks}

We have compared the post-shock electron and proton temperatures in
\dem\ to determine the initial electron-ion equilibration, $g_{0}
\equiv (T_{\rm e}/T_{\rm p})_{0}$, over a range of shock
speeds. First, we conducted independent measurements of \te from
spatially resolved \chandra\ ACIS-S X-ray spectra of the outer rim,
using three different nonequilibrium ionization models for the
post-shock temperature evolution.  The first two models assumed a
constant velocity shock, one with full initial equilibration, $g_{0} =
1$, the other with only minimal initial equilibration, \gmemp ,
followed by post-shock Coulomb collisional equilibration. The third
model was a fully equilibrated ($g_{0} = 1$) NEI Sedov solution.  In
all cases the ionization state was allowed to progress downstream.  We
compared the inferred average shock temperatures with the proton
temperatures from the H$\alpha$ line profiles reported in GRHW03.  The
slowest shock was consistent with full equilibration in both the
constant velocity and Sedov models, but not with minimal heating. In
contrast all the faster shocks were consistent with \gmemp but not
$g_{0} = 1$.

Our independent measurements of \te and \tp showed that all regions
were consistent with a planar shock model, where the electron and
proton temperatures started at some initial degree of equilibration,
$g_{0}$, and then slowly equilibrated downstream due to Coulomb
collisions.  Given this agreement, we constructed a planar X-ray
model which incorporated the measured value of $T_{\rm p}$ and allowed
for a variable $g_{0}$.  With the additional constraints from the
proton temperature we found one region, X4, with intermediate shock
velocity, for which both full and minimal equilibration could be ruled
out, albeit only at 1$\sigma$.  Our results support the hypothesis
from Ghavamian et~al.~(2001), that there is an anti-correlation
between shock speed and initial equilibration.

The method presented here, of comparing the optically determined
$T_{\rm p}$ to the X-ray measured $T_{\rm e}$, is a powerful new
technique for constraining the initial temperature equilibration of
SNR shock fronts.  Its strength lies in the fact that neither the
proton nor the electron temperature are strongly model dependent, as
this work shows.  On the other hand, \dem\ was a particularly good
remnant for this study. We do not expect cosmic-ray emission to
contaminate our blast wave spectra because \dem\ has the lowest radio
flux of any of the 25 LMC SNRs in Mathewson et~al.~(1983) and there is
no evidence in the X-rays for a high energy continuum from synchrotron
emission (Hughes et al.~2003).  Additionally, the clear separation
between the ejecta and the blast wave allowed us to obtain blast wave
spectra that are uncontaminated by SN ejecta. Other Balmer-dominated
SNRs, such as Tycho or SN1006, may suffer more from an intermingling
of ejecta into the blast wave zone, and complications to the fit from
a cosmic ray synchrotron component.

All the shocks we studied in \dem\ are in the interesting velocity
range where intermediate initial equilibrations appear to
occur. However, due to the large errors on our current measurements of
$g_{0}$, only one region was shown to be of intermediate
equilibration, although two other regions were best fit by
intermediate values of $g_{0}$. In the current fits these errors were
dominated by the fitted error on the electron temperature, by the
root-mean-square error on the abundances and column density, and to a
lesser extent by the error in the proton temperature.  We have been
awarded a re-observation of \dem\ with {\it Chandra} to quadruple the
total exposure time. With this observation we will constrain the
abundances of all elements with prominent lines separately for each
aperture and obtain tighter limits on the electron temperature.  The
increased statistics will also allow us to constrain the temperatures
and timescales of each nested region separately. Hence we will be able
to directly determine the temperature variation downstream of the
shock, and verify whether it follows the evolution given by our shock
models.  Additional optical spectroscopy of the non-radiative shock
regions identified with the Fabry-Perot would be useful to reduce the
uncertainties in the proton temperatures, and better sample the broad
H$\alpha$ component.

\begin{acknowledgements}
We are grateful for the use of Leisa Townsley's CTI-correction
algorithm and associated response matrices.  We are pleased to
acknowledge useful discussions with A. Sluis and
T. B. Williams. Partial support was provided by NASA/Chandra grants
GO0-1035X, G01-2052X, and G02-3068X; CER was supported by a NASA
Graduate Student Researchers Program Fellowship.

\end{acknowledgements}

\clearpage

\figcaption[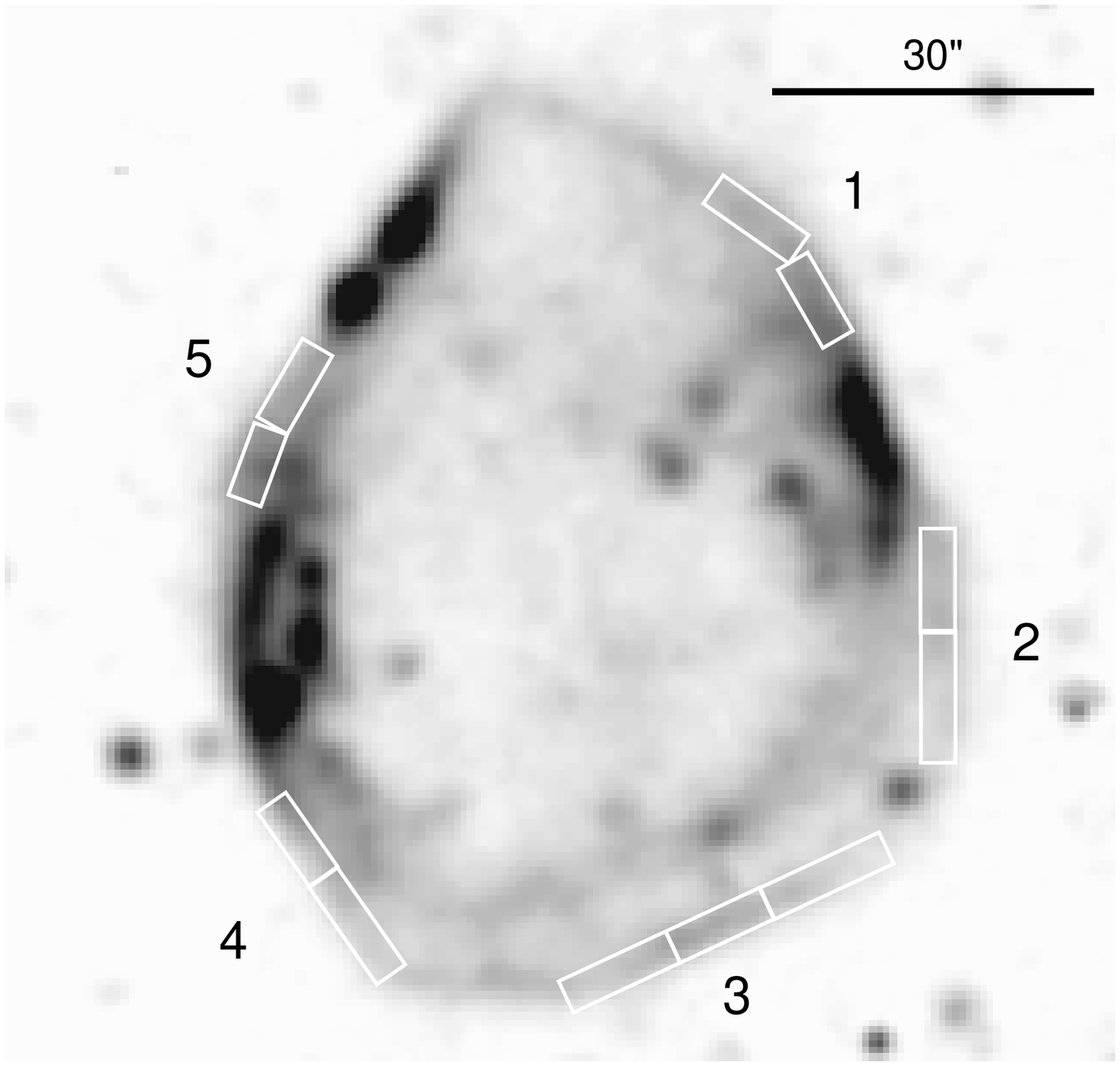]{
\label{fig:halphaimage} H$\alpha$ image of DEM~L71. Rutgers Fabry-Perot 
(RFP) image of supernova remnant (SNR) DEM~L71 at the wavelength of
the narrow H$\alpha$ line.  Blast wave regions from which we extracted
spectra are as marked.}

\figcaption[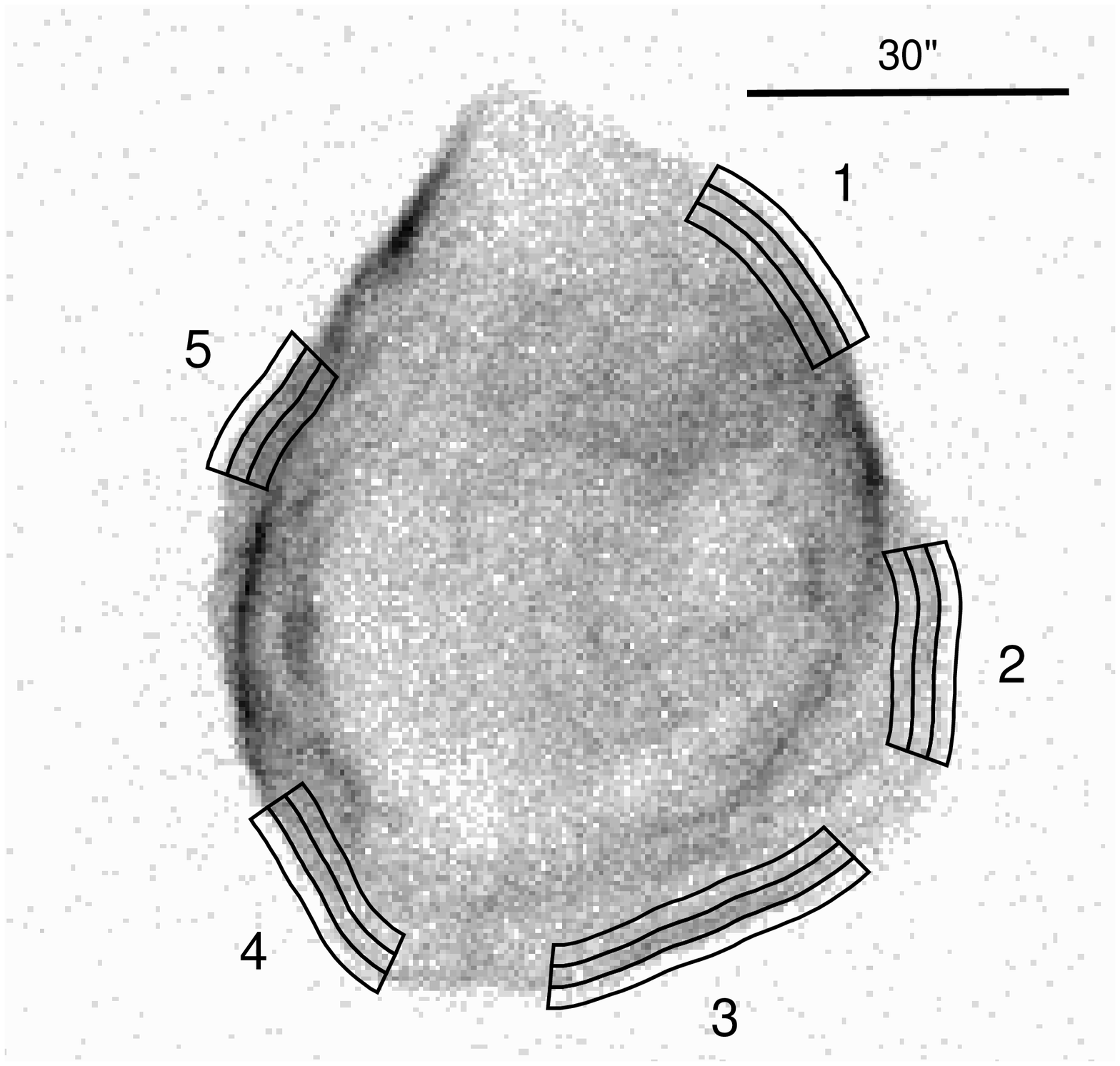]{
\label{fig:chandraimage} {\it Chandra} soft X-ray image of DEM~L71 
(0.2$-$0.75~keV). X-ray spectra were extracted from three consecutive
regions behind the shock at each aperture. }

\figcaption[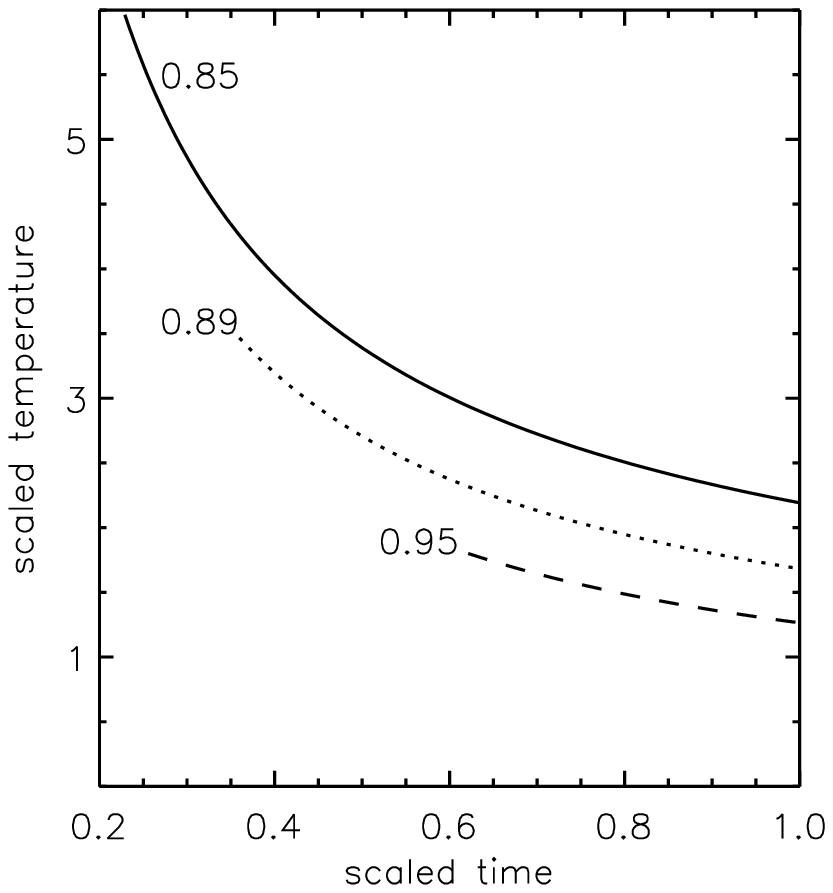]{
\label{fig:sedovtemp} Sedov temperature evolution. The Sedov 
solution for the ratio, $T_{ave}/T_{ave,s}$, of the average
temperature of a parcel of gas to the current average shock
temperature as a function of the ratio, $t / t_{c}$, of the evolving
time to the current time. The evolution is plotted for the innermost
parcel of material in each of the three regions behind the shock,
labeled by their current fractional radii. Note that for the $g_{0}=1$
case $T_{\rm e}=T_{ave}$.}

\figcaption[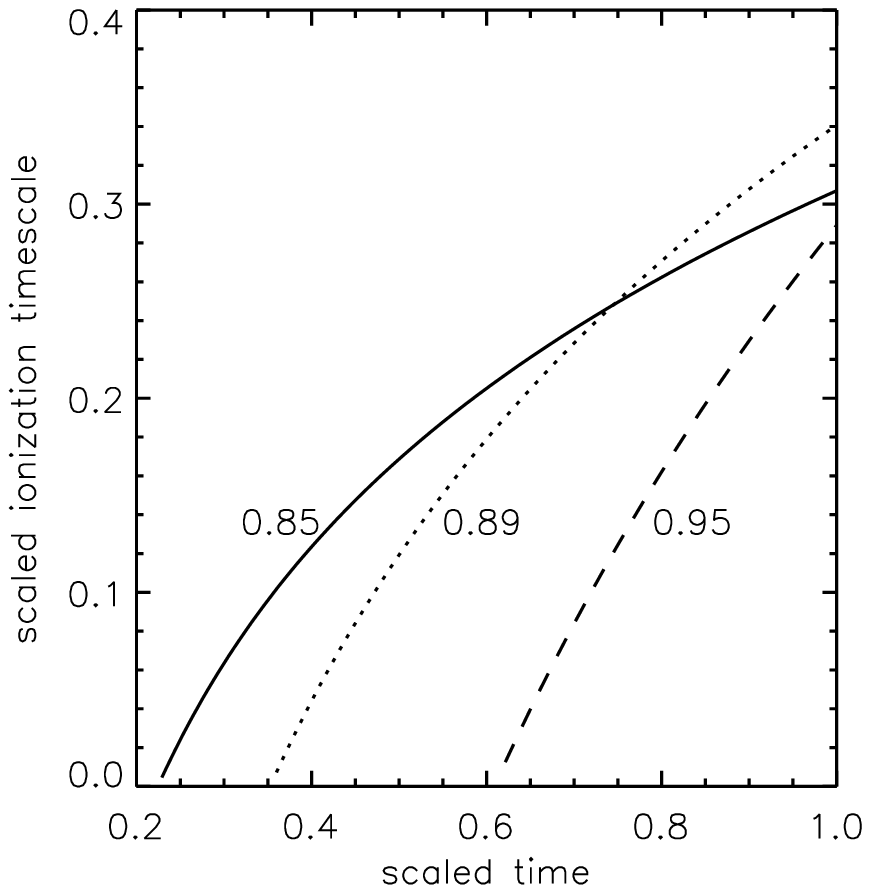]{
\label{fig:sedovnet} Sedov ionization timescale evolution.  The Sedov 
solution for the ratio of the evolving ionization timescale, $n_{\rm
e}t$, to the product of the ambient electron density with the current
age of the remnant.  This is plotted as a function of $t / t_{c}$ for
the innermost parcel of material in each of the three regions behind
the shock, labeled by their current fractional radii. }

\figcaption[fig5abcde]{
\label{fig:spectra} X-ray spectra of the blast wave. For each aperture 
we plot the {\it Chandra} X-ray spectra from three consecutive regions
behind the shock, outermost to innermost, bottom to top. The model
shown is a non-equilibrium ionization (NEI) planar shock with minimal
initial temperature equilibration, $g_{0} = m_{\rm e}/m_{\rm p}$.  }

%\clearpage

\figcaption[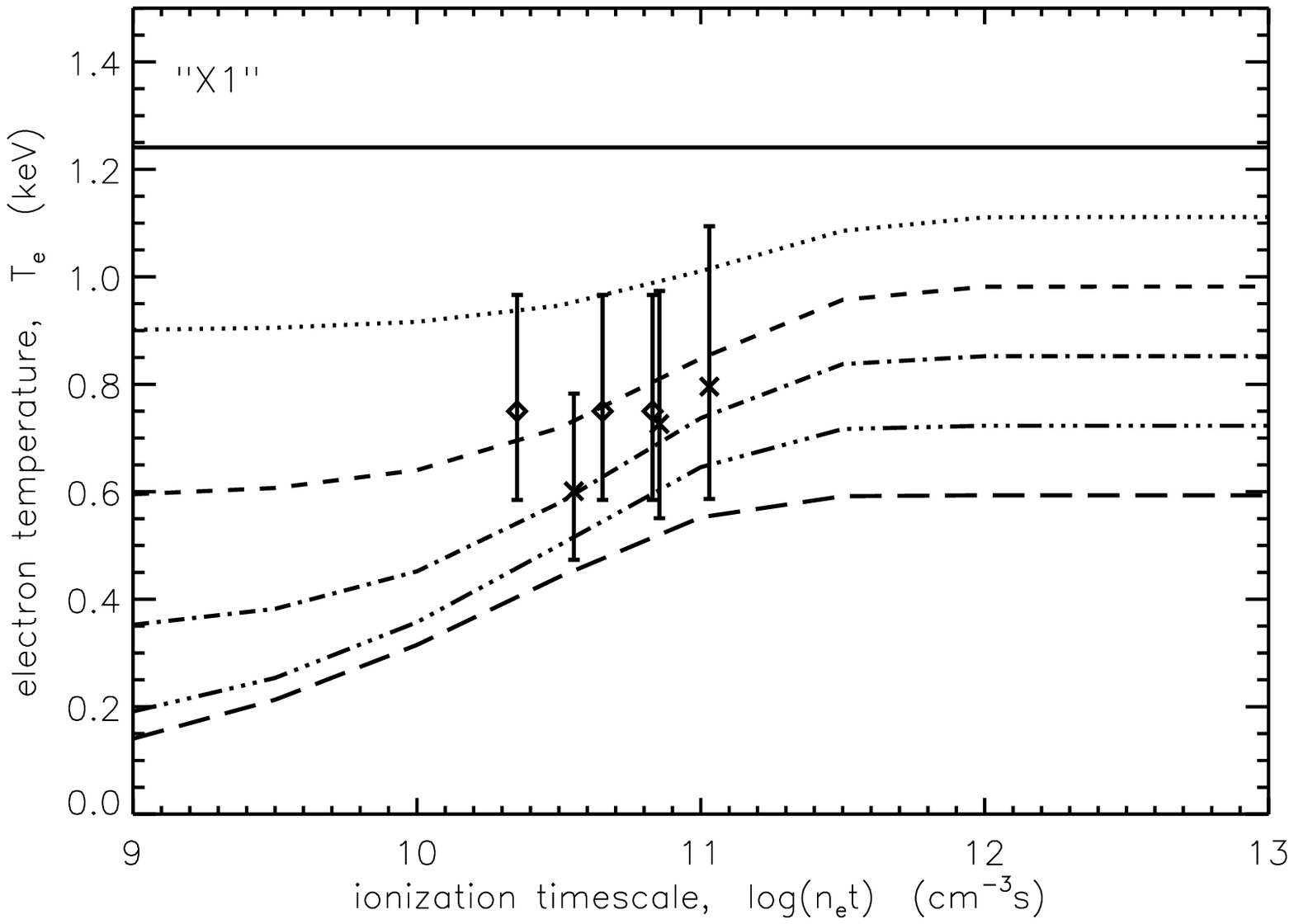]{
\label{fig:evolve} Temperature evolution behind a planar shock.
Comparison between the $T_{\rm p}$ measured optically and the $T_{\rm
e}$ measured in the X-rays for aperture X1. The data points represent
$T_{\rm e}$ for three consecutive regions behind the shock for the
best-fit minimally equilibrated ($g_{0} = m_{\rm e}/m_{\rm p} $ ),
crosses, and fully equilibrated ($g_{0} = 1$), diamonds, planar
shocks.  In both cases, the three data points were not measured
independently but rather were fitted together for a single average
temperature.  The error bars include the 1$\sigma$ errors on $T_{\rm
e}$ as well as the RMS errors on the abundances and column
density. The curves represent $T_{\rm e}$ as a function of ionization
timescale implied by the optically determined $T_{\rm p}$ for various
$g_{0}$: 1.0, 0.8, 0.6, 0.4, 0.2, $m_{\rm e}/m_{\rm p}$, from solid to
dashed.  }

\figcaption[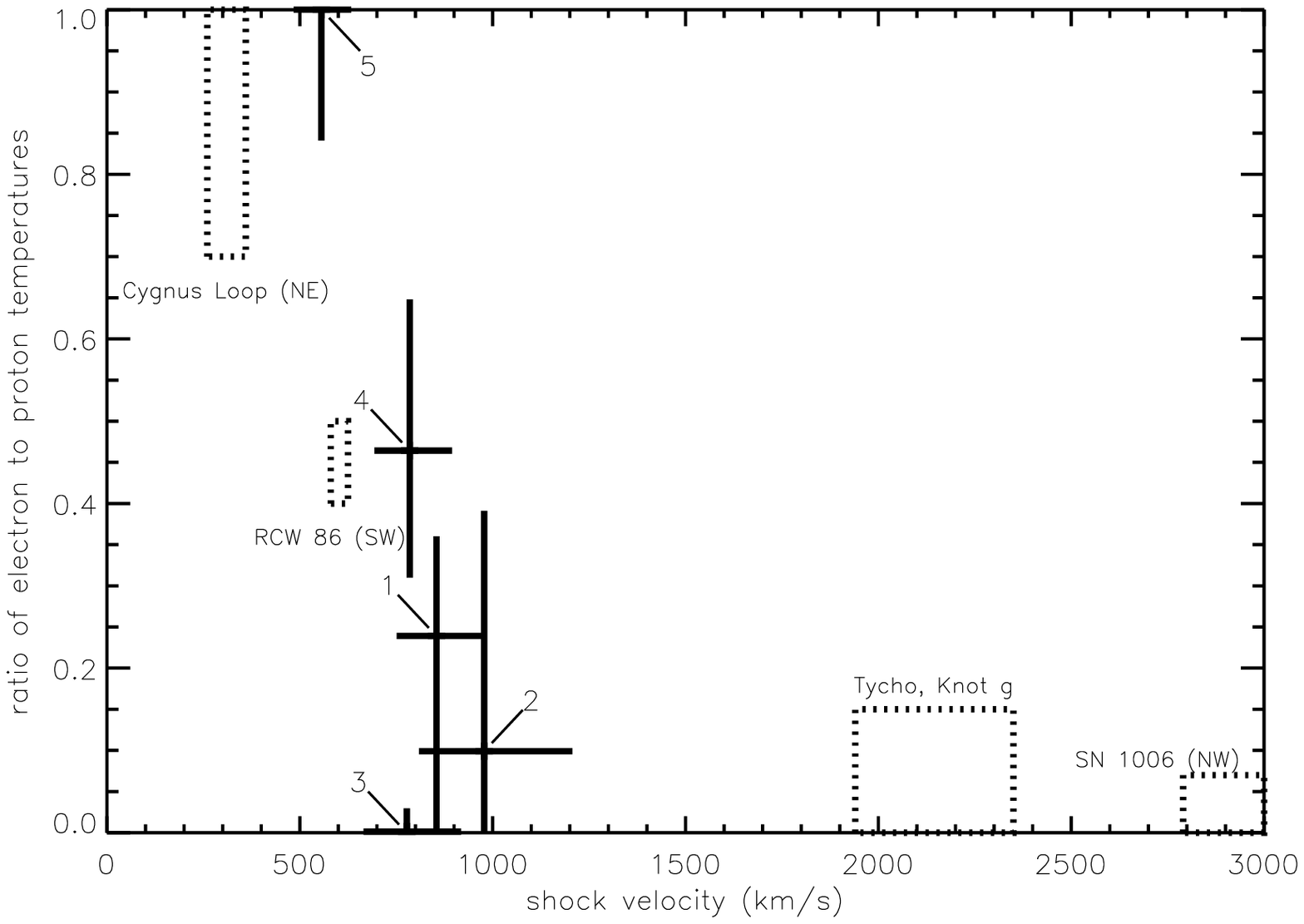]{
\label{fig:v_vs_g} The degree of initial electron-proton equilibration. 
The ratio of the initial electron to proton temperatures, $g_{0}$, is
plotted as a function of shock speed for the five regions in DEM~L71
(solid error bars). These are consistent with the anti-correlation
between $g_{0}$ and shock speed found from modeling the
broad-to-narrow flux ratios for the Balmer filaments in other remnants
(dotted boxes, Ghavamian et~al.~2001). The uncertainties in the
equilibration for the DEM~L71 regions include the 1$\sigma$ error bars
on the best-fit $g_{0}$ at the nominal $T_{\rm p}$, the range in
$g_{0}$ within the 1$\sigma$ error bars on $T_{\rm p}$, and the range
in $g_{0}$ within the RMS errors on the abundance and column
density, all summed in quadrature. The range of velocities comes from
the 1$\sigma$ range in proton temperatures at the best-fit
$g_{0}$. For the other remnants the boxes represent the allowed range
of velocities and equilibrations for which both the width of the broad
line and the broad-to-narrow ratio could be modeled.  }

%%%UCP%%%
%\newpage
%\plotone{fig1.eps}
%\newpage
%\plotone{fig2.eps}
%\newpage
%\plotone{fig3.eps}
%\newpage
%\plotone{fig4.eps}
%\newpage
%\plotone{fig5a.eps}
%\newpage
%\plotone{fig5b.eps}
%\newpage
%\plotone{fig5c.eps}
%\newpage
%\plotone{fig5d.eps}
%\newpage
%\plotone{fig5e.eps}
%\newpage
%\plotone{fig6.eps}
%\newpage
%\plotone{fig7.eps}

\clearpage

\begin{deluxetable}{cc}
\tablewidth{0pt}
\tablecaption{\label{ta:apertures} }
\tablehead{ RFP apertures & \chandra\   apertures }
\startdata
3, 4 & X1 \\
5, 6 & X2 \\
7, 8, 9 & X3 \\
11, 12 & X4 \\
13, 14 & X5 \\
\enddata
\end{deluxetable}

\begin{deluxetable}{ccccccc}
\tablewidth{0pt}
\tablecaption{\label{ta:RFP} RFP H$\alpha$ Results for Selected Regions in the DEM L 71 Blast Wave }
\tablehead{
	& \multicolumn{2}{c}{H$\alpha$ profile fits} 
	& \multicolumn{2}{c}{$(T_{\rm e}/T_{\rm p})_{0} = 1$} 
	& \multicolumn{2}{c}{$(T_{\rm e}/T_{\rm p})_{0} = m_{\rm e}/m_{\rm p}$} 
}
\startdata
	Aperture & V$_{FWHM}$ (km s$^{-1}$) & I$_{B}$/I$_{N}$ & 
	V$_{shock}$ (km s$^{-1}$) & $kT_{\rm p}$ (keV) &
	V$_{shock}$ (km s$^{-1}$) & $kT_{\rm p}$ (keV) \\  \hline
1 & 840$^{+115}_{-100}$ & 0.51$^{+0.06}_{-0.06}$  &
	1050$^{+140}_{-130}$ & 1.29$^{+0.38}_{-0.30}$ &
	815$^{+115}_{-100}$ & 1.29$^{+0.38}_{-0.30}$ \\
2 & 985$^{+210}_{-165}$ & 0.54$^{+0.09}_{-0.09}$  &
	1240$^{+290}_{-210}$ & 1.81$^{+0.95}_{-0.56}$ &
	960$^{+215}_{-165}$ & 1.81$^{+0.90}_{-0.57}$ \\
3 & 805$^{+140}_{-115}$ & 0.49$^{+0.07}_{-0.06}$  &
	1005$^{+170}_{-150}$ & 1.18$^{+0.44}_{-0.32}$ & 
	775$^{+140}_{-115}$ & 1.18$^{+0.46}_{-0.32}$ \\
4 & 735$^{+100}_{-85}$ & 0.66$^{+0.08}_{-0.08}$  & 
	915$^{+130}_{-105}$ & 0.99$^{+0.29}_{-0.21}$ & 
	710$^{+100}_{-80}$ & 0.98$^{+0.29}_{-0.21}$ \\
5 & 450$^{+60}_{-60}$ & 0.44$^{+0.06}_{-0.05}$  &
	555$^{+75}_{-70}$ & 0.36$^{+0.10}_{-0.09}$ & 
	430$^{+60}_{-55}$ & 0.36$^{+0.11}_{-0.09}$  \\
\enddata
\end{deluxetable}

\begin{deluxetable}{ccccc}
\tablewidth{0pt}
\tablecaption{\label{abundances}Average $N_{\mathrm H}$ and abundances}
\tablehead{
model & $N_{\mathrm H}$ & O & Ne & Fe \\
 & 10$^{20}$ atoms cm$^{-2}$ & dex & dex & dex 
}
\startdata
$g_{0}=1$, planar & 5.8$\pm$1.8 & 8.39$_{-0.12}^{+0.09}$ & 7.64$_{-0.15}^{+0.11}$ & 6.86$_{-0.11}^{+0.09}$ \\
$g_{0}= \frac{m_{\rm e}}{m_{\rm p}}$, planar & 6.0$\pm$1.7 & 8.40$_{-0.13}^{+0.10}$ & 7.66$_{-0.15}^{+0.11}$ & 6.86$_{-0.14}^{+0.11}$ \\
$g_{0}=1$, Sedov & 7.1$\pm$2.3 & 8.41$_{-0.13}^{+0.10}$ & 7.78$_{-0.08}^{+0.07}$ & 6.83$_{-0.12}^{+0.09}$ \\
combined & 6.3$\pm$1.9 & 8.40$_{-0.12}^{+0.09}$ & 7.70$_{-0.14}^{+0.10}$ & 6.86$_{-0.11}^{+0.09}$ 
\enddata
\end{deluxetable}

\begin{deluxetable}{cccccc}
\tablewidth{0pt}
\tablecaption{\label{xrayparameters} $Chandra$ X-ray Best-fit Model Parameters for Selected Regions in the DEM~L71 Blast Wave}
\tablehead{
	Aperture & NEI model & $kT_{ave,s}$ (keV) 
	& log($n_{\rm e}t_{final}$) cm$^3$s 
	& $\chi ^{2}$ & r-$\chi ^{2}$
}
\startdata
1 & $g_{0}=1$, planar & 0.75$^{+0.22}_{-0.17}$ & 10.83$^{+0.11}_{-0.20}$ 
	& 129.45 & 1.455 \\
1 & $g_{0}= \frac{m_{\rm e}}{m_{\rm p}}$, planar 
	& 0.95$^{+0.62}_{-0.32}$ & 11.03$^{+0.10}_{-0.18}$ 
	& 124.78 & 1.402 \\
1 & $g_{0}=1$, Sedov & 0.57$^{+0.12}_{-0.13}$ & 10.34$^{+0.09}_{-0.13}$ 
	& 143.93 & 1.617 \\  \hline
2 & $g_{0}=1$, planar & 0.94$^{+0.30}_{-0.39}$ & 10.68$^{+0.18}_{-0.30}$ 
	& 85.84 & 1.431 \\
2 & $g_{0}= \frac{m_{\rm e}}{m_{\rm p}}$, planar 
	& 1.26$^{+0.79}_{-0.72}$ & 10.97$^{+0.15}_{-0.23}$ 
	& 79.76 & 1.323 \\
2 & $g_{0}=1$, Sedov & 0.74$^{+0.43}_{-0.36}$ &  10.23$^{+0.14}_{-0.34}$ 
	& 82.79 & 1.380 \\  \hline
3 & $g_{0}=1$, planar & 0.66$^{+0.05}_{-0.12}$ & 11.27$^{+0.10}_{-0.12}$ 
	& 158.69 & 1.430 \\
3 & $g_{0}= \frac{m_{\rm e}}{m_{\rm p}}$, planar 
	& 0.70$^{+0.08}_{-0.11}$ & 11.42$^{+0.06}_{-0.06}$ 
	& 144.36 & 1.301 \\
3 & $g_{0}=1$, Sedov & 0.34$^{+0.05}_{-0.01}$ &  11.13$^{+0.15}_{-0.06}$ 
	& 174.74 & 1.574 \\  \hline
4 & $g_{0}=1$, planar & 0.72$^{+0.20}_{-0.16}$ & 10.95$^{+0.11}_{-0.18}$ 
	& 73.54 & 0.994 \\
4 & $g_{0}= \frac{m_{\rm e}}{m_{\rm p}}$, planar 
	& 0.94$^{+0.70}_{-0.42}$ & 11.11$^{+0.14}_{-0.23}$ 
	& 78.15 & 1.056 \\
4 & $g_{0}=1$, Sedov & 0.66$^{+0.16}_{-0.22}$ & 10.37$^{+0.09}_{-0.18}$ 
	& 76.25 & 1.030 \\  \hline
5 & $g_{0}=1$, planar & 0.416$^{+0.033}_{-0.036}$ & 11.9$^{+0.08}_{-0.12}$ 
	& 164.85 & 1.459 \\
5 & $g_{0}= \frac{m_{\rm e}}{m_{\rm p}}$, planar 
	& 0.42$^{+0.04}_{-0.03}$ & 11.93$^{+0.08}_{-0.09}$ 
	& 159.37 & 1.410 \\
5 & $g_{0}=1$, Sedov & 0.25$^{+0.02}_{-0.02}$ &  11.72$^{+0.17}_{-0.77}$ 
	& 152.21 & 1.348 \\
\enddata
\end{deluxetable}

\begin{deluxetable}{cccccccc}
\tablewidth{0pt}
%\tablecaption{\label{3regktnt} $Chandra$ X-ray Results for Selected Regions in the DEM~L71 Blast Wave}
\tablecaption{\label{3regktnt} Variation in $kT_{\rm e}$ and $n_{\rm e}t$ across the 3 Nested Regions Implied by each Shock Model}
\tablehead{
	& & \multicolumn{2}{c}{outer region} & \multicolumn{2}{c}{middle region} & \multicolumn{2}{c}{inner region} \\
	Aperture & NEI model
	& $kT_{\rm e}$  & log($n_{\rm e}t$)  
	& $kT_{\rm e}$  & log($n_{\rm e}t$)  
	& $kT_{\rm e}$  & log($n_{\rm e}t$)  \\
	& & keV & cm$^3$s & keV & cm$^3$s &  keV & cm$^3$s 
}
\startdata
1 & $g_{0}=1$, planar & 
	0.75  & 10.35
	& 0.75  & 10.65 
	& 0.75  & 10.83  \\
1 & $g_{0}= \frac{m_{\rm e}}{m_{\rm p}}$, planar 
	& 0.60 & 10.55 
	& 0.73 & 10.85 
	& 0.80 & 11.03 \\
1 & $g_{0}=1$, Sedov & 
	0.72 & 10.31 
	& 0.96  & 10.38 
	& 1.25  & 10.34 \\  \hline
2 & $g_{0}=1$, planar &
	0.94 & 10.20
	& 0.94 & 10.50 
	& 0.94 & 10.68 \\
2 & $g_{0}= \frac{m_{\rm e}}{m_{\rm p}}$, planar 
	& 0.67 & 10.49 
	& 0.83 & 10.79
	& 0.93 & 10.97 \\
2 & $g_{0}=1$, Sedov & 
	0.93 & 10.20
	& 1.23 & 10.27 
	& 1.60 & 10.23 \\  \hline
3 & $g_{0}=1$, planar &
	0.66 & 10.79
	& 0.66 & 11.09
	& 0.66 & 11.27 \\
3 & $g_{0}= \frac{m_{\rm e}}{m_{\rm p}}$, planar 
	& 0.62 & 10.94 
	& 0.67 & 11.24
	& 0.69 & 11.42 \\
3 & $g_{0}=1$, Sedov & 
	0.42 & 11.11
	& 0.56 & 11.18
	& 0.73 & 11.14 \\  \hline
4 & $g_{0}=1$, planar &
	0.72 & 10.47
	& 0.72 & 10.77
	& 0.72 & 10.95 \\
4 & $g_{0}= \frac{m_{\rm e}}{m_{\rm p}}$, planar 
	& 0.63 & 10.63 
	& 0.75 & 10.93
	& 0.82 & 11.11 \\
4 & $g_{0}=1$, Sedov & 
	0.82 & 10.34 
	& 1.09  & 10.41 
	& 1.42  & 10.37  \\  \hline
5 & $g_{0}=1$, planar &
	0.416 & 11.42
	& 0.416 & 11.72
	& 0.416 & 11.90 \\
5 & $g_{0}= \frac{m_{\rm e}}{m_{\rm p}}$, planar 
	& 0.42 & 11.45 
	& 0.42 &  11.75
	& 0.42 & 11.93 \\
5 & $g_{0}=1$, Sedov & 
	0.31 & 11.69
	& 0.41 & 11.76
	& 0.53 & 11.71 \\  \hline
\enddata
\end{deluxetable}

\begin{deluxetable}{ccccc}
\tablewidth{0pt}
\tablecaption{\label{gzero} Initial Electron-Ion Equilibration }
\tablehead{
	& \multicolumn{1}{c}{RFP H$\alpha$} 
	& \multicolumn{2}{c}{$Chandra$} & \\
	& & \multicolumn{2}{c}{outermost region} & 
}
\startdata
	Aperture   & $kT_{\rm p}$ 
	& $kT_{\rm e}$ & log($n_{\rm e}t$) & 
	$g_{0} \equiv (T_{\rm e}/T_{\rm p})_{0}$ \\  
	 & (keV) & (keV) & cm$^3$s & \\  \hline
1 & $1.29^{+0.38}_{-0.30}$
	& 0.60$^{+0.18}_{-0.12}$ & 10.55$^{+0.10}_{-0.18}$ 
	& 0.41$^{+0.31}_{-0.41}$ \\
2 & 1.81$^{+0.95}_{-0.57}$
	& 0.67$^{+0.18}_{-0.25}$ & 10.49$^{+0.15}_{-0.23}$ 
	& 0.33$^{+0.32}_{-0.33}$ \\
3 & 1.18$^{+0.46}_{-0.32}$
	& 0.62$^{+0.05}_{-0.08}$ & 10.94$^{+0.06}_{-0.06}$ 
	& 0.26$^{+0.39}_{-0.26}$ \\
4 & 0.98$^{+0.29}_{-0.21}$
	& 0.63$^{+0.22}_{-0.19}$ & 10.63$^{+0.14}_{-0.23}$ 
	& 0.64$^{+0.36}_{-0.54}$ \\
5 & 0.36$^{+0.11}_{-0.09}$
	& 0.42$^{+0.04}_{-0.03}$ & 11.45$^{+0.08}_{-0.09}$ 
	& 1.0$_{-0.13}$ \\
\enddata
\end{deluxetable}

\begin{deluxetable}{cccc}
\tablewidth{0pt}
\tablecaption{\label{gzerofree} Variable $g_{0}$ Model}
\tablehead{
	Aperture & $V_{s}$ (km s$^{-1}$) & $kT_{\rm p}$ (keV) & $g_{0}$
}
\startdata
1 & 855$^{+120}_{-105}$ & $1.29^{+0.38}_{-0.30}$ & 0.24$^{+0.12}$ \\
2 & 980$^{+230}_{-170}$ & 1.81$^{+0.95}_{-0.57}$ & 0.10$^{+0.29}$ \\
3 & 775$^{+140}_{-110}$ & 1.18$^{+0.46}_{-0.32}$ & 0.01$^{+0.03}$ \\
4 & 785$^{+110}_{-90}$ & 0.98$^{+0.29}_{-0.21}$ & 0.46$^{+0.18}_{-0.15}$ \\
5 & 555$^{+75}_{-70}$ & 0.36$^{+0.11}_{-0.09}$ & 1.0$_{-0.16}$ \\
\enddata
\end{deluxetable}

%\end{document}

\clearpage

\begin{figure}
\epsfig{figure=fig1.eps, width=4.0in}
\end{figure}
\begin{figure}
\epsfig{figure=fig2.eps, width=4.0in}
\end{figure}

\clearpage

\begin{figure}
\epsfig{figure=fig3.eps}
\end{figure}
\begin{figure}
\epsfig{figure=fig4.eps}
\end{figure}

\clearpage

\begin{figure}
\epsfig{figure=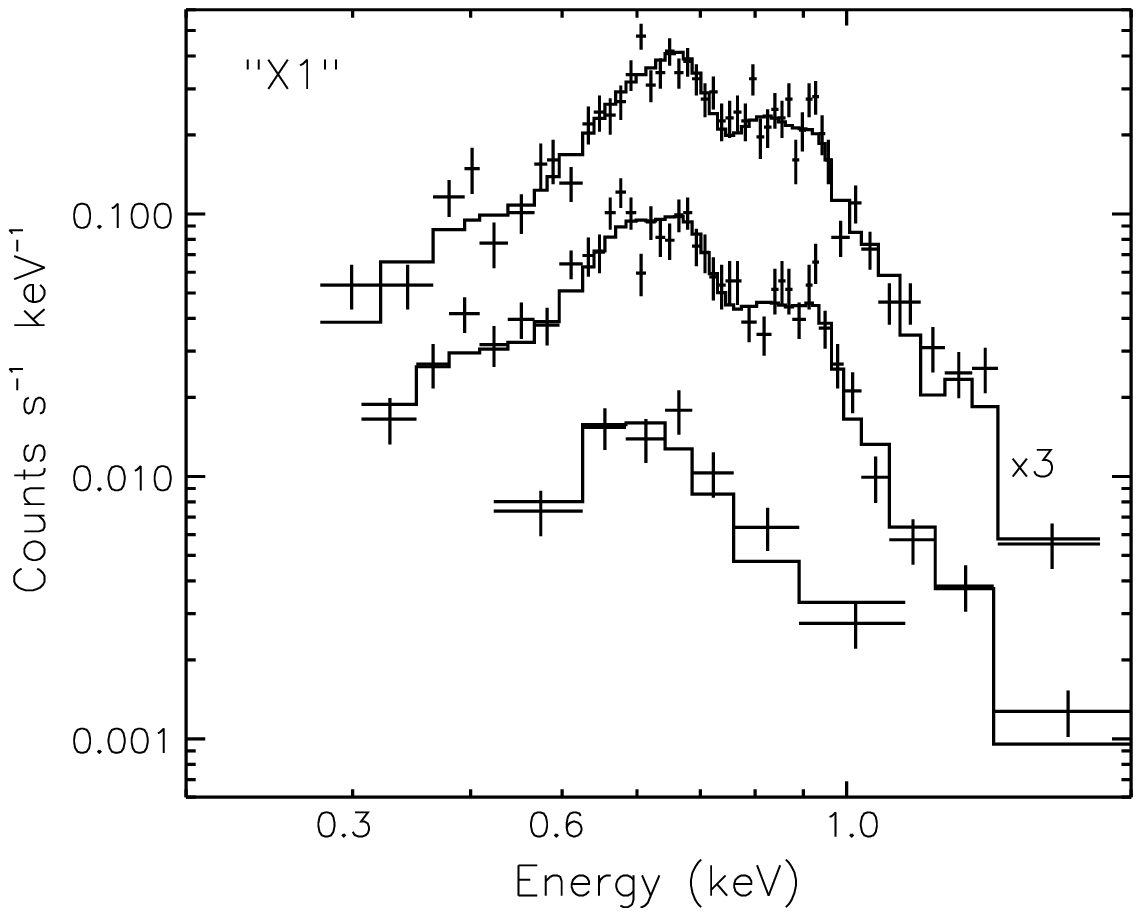}
\end{figure}
\begin{figure}
\epsfig{figure=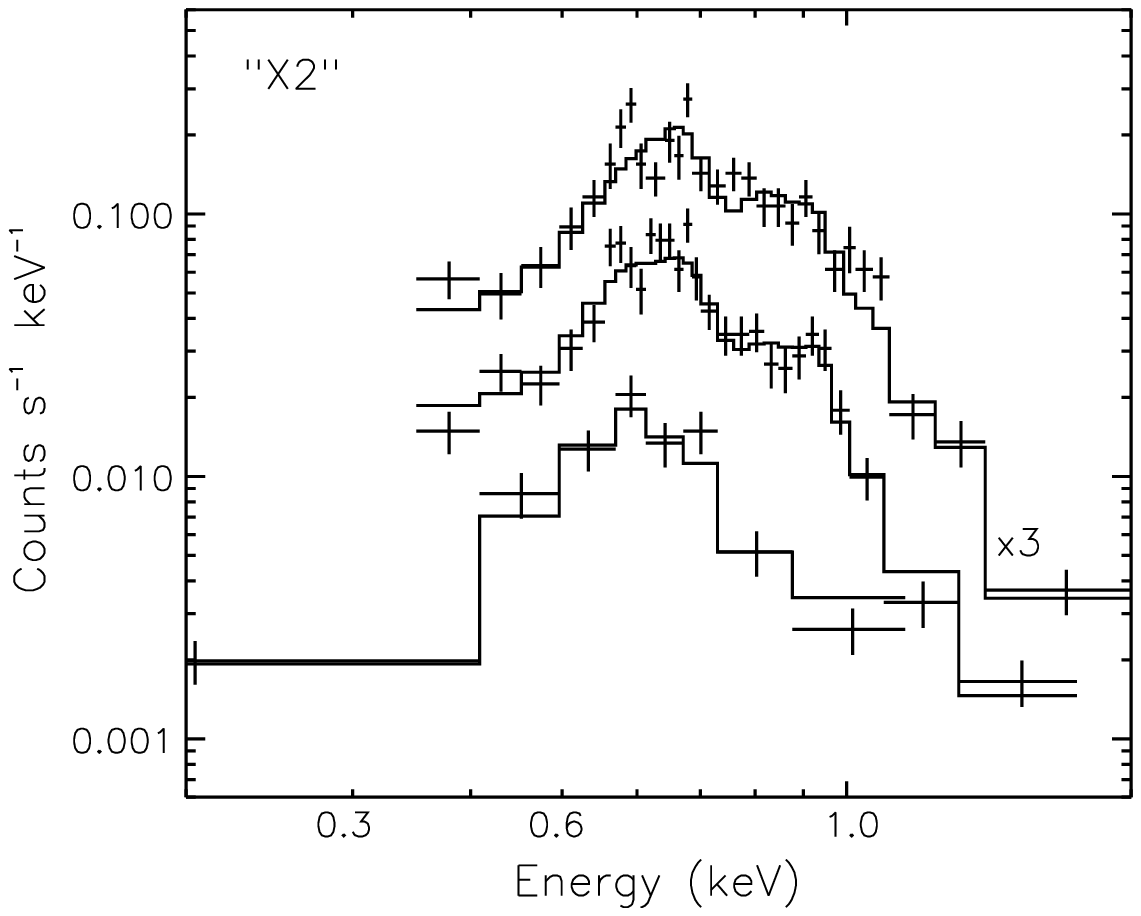}
\end{figure}

\clearpage

\begin{figure}
\epsfig{figure=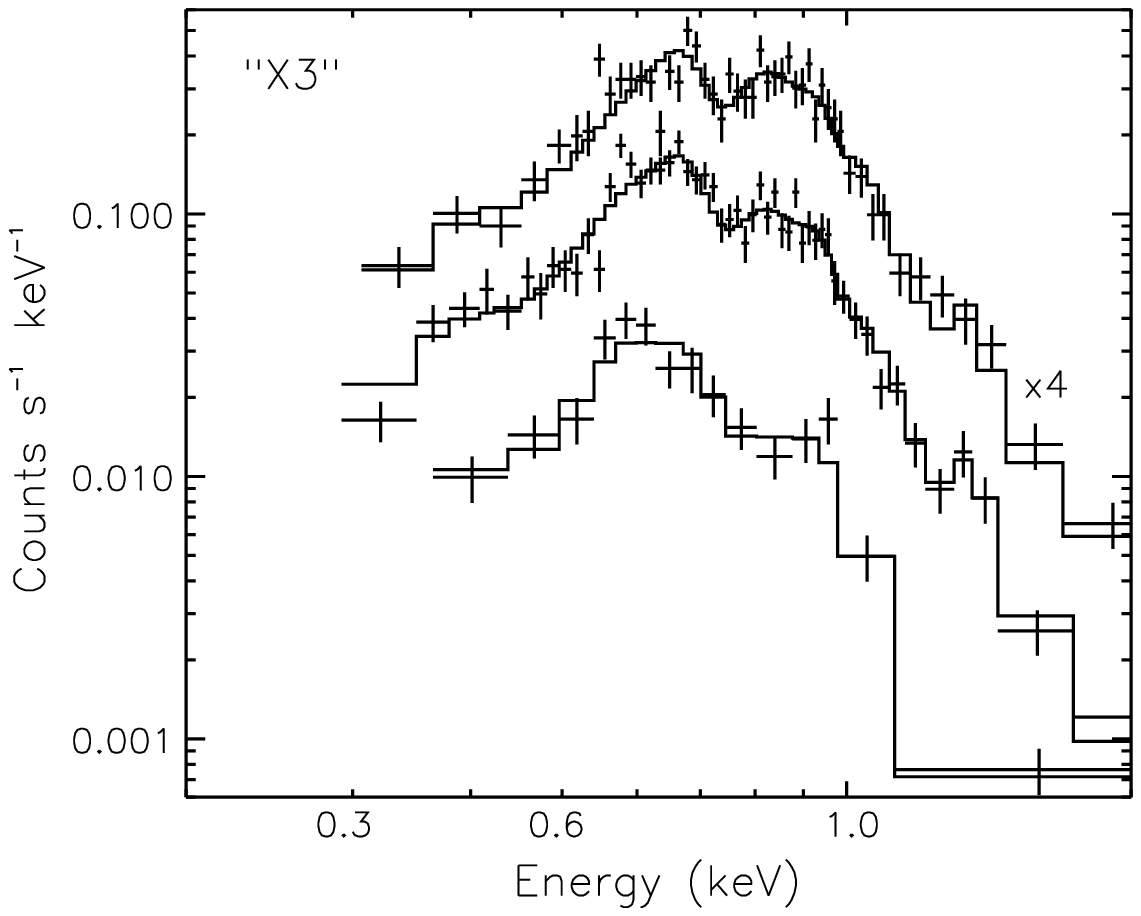}
\end{figure}
\begin{figure}
\epsfig{figure=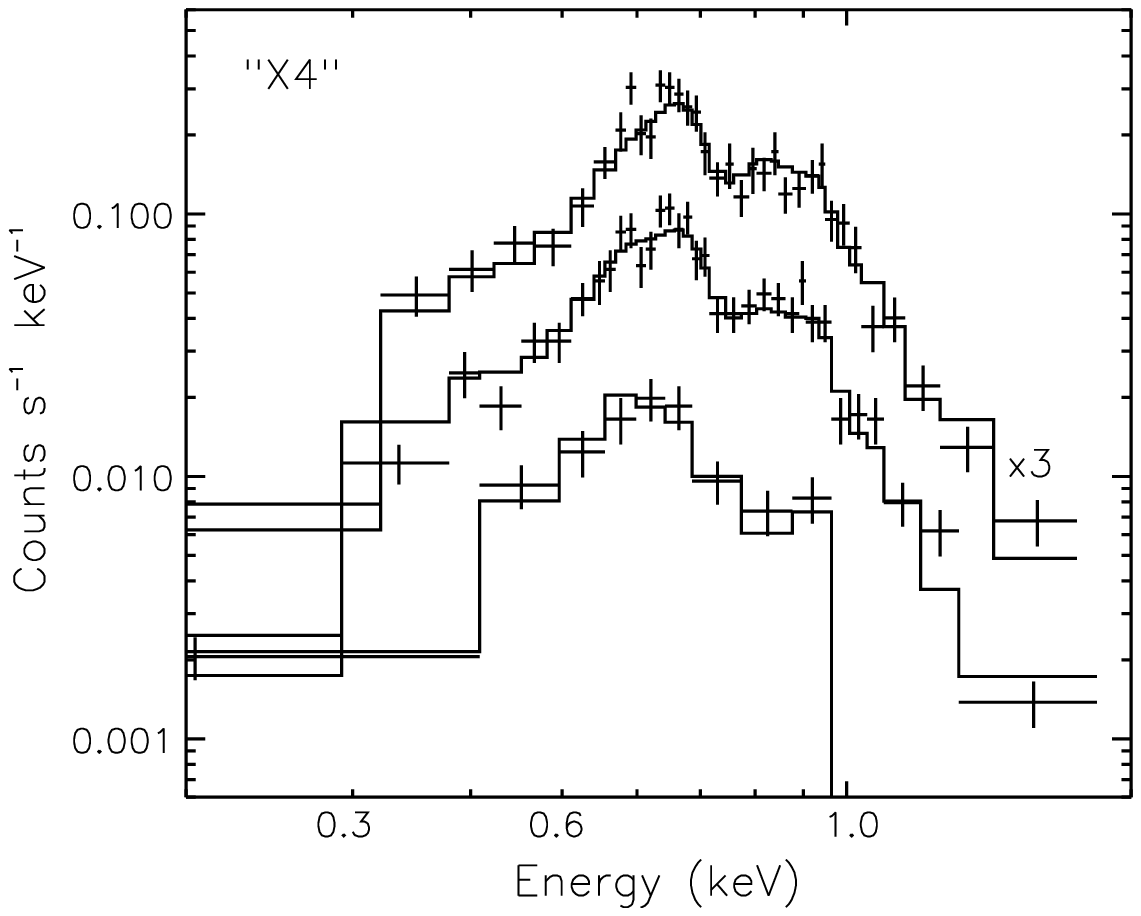}
\end{figure}

\clearpage

\begin{figure}
\epsfig{figure=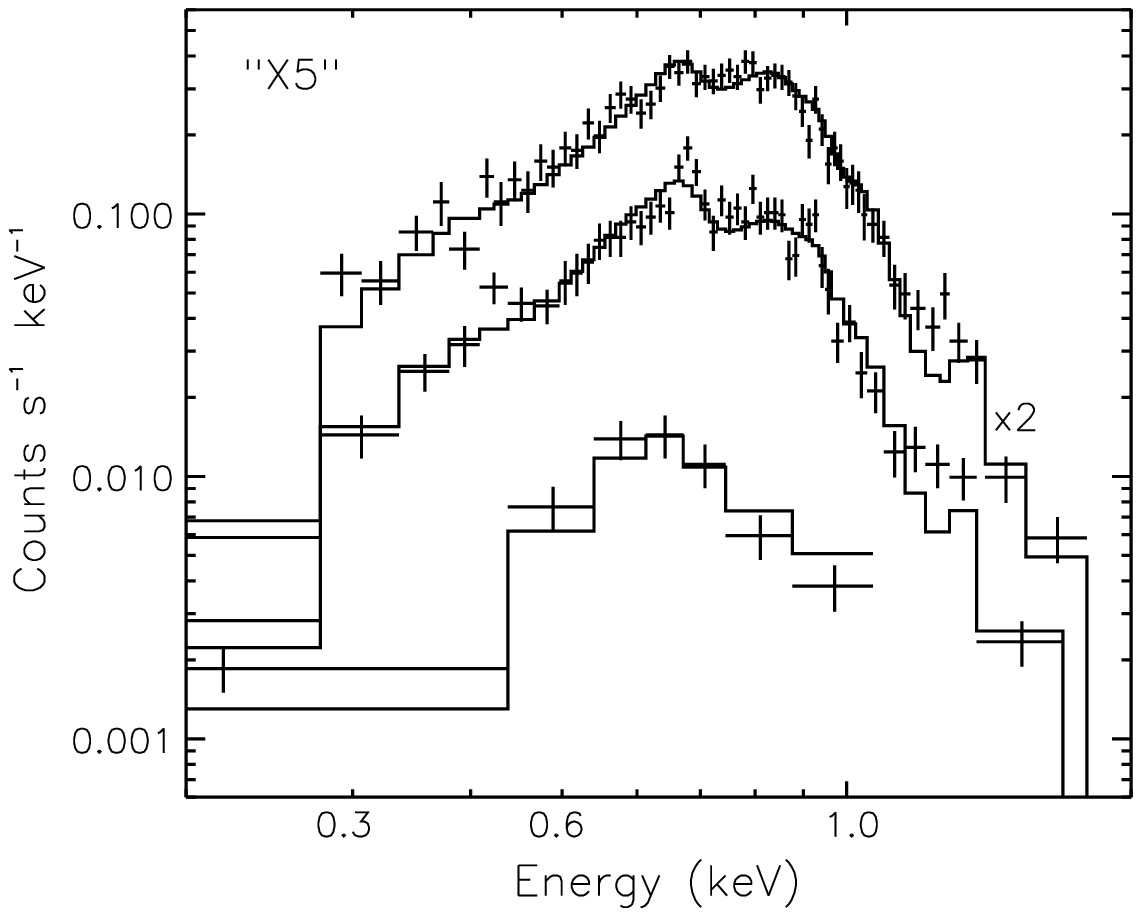}
\end{figure}

\clearpage

\begin{figure}
\epsfig{figure=fig6.eps}
\end{figure}

\clearpage

\begin{figure}
\epsfig{figure=fig7.eps}
\end{figure}

\end{document}